\documentclass[prl,twocolumn]{revtex4}
\usepackage{graphicx}
\usepackage{dcolumn}
\usepackage{amsmath,amsbsy,amssymb,graphics}
\usepackage{bm}
\usepackage{mathrsfs}

\begin{document}

\preprint{APS/123-QED}

\title{
Superconductivity of Composite Particles in Two-Channel Kondo Lattice
}

\author{Shintaro Hoshino$^1$ and Yoshio Kuramoto$^2$}

\affiliation{
$^1$Department of Basic Science, The University of Tokyo, Meguro, Tokyo 153-8902, Japan
\\
$^2$Department of Physics, Tohoku University, Sendai, Miyagi 980-8578, Japan
}

\date{\today}

\begin{abstract}
Emergence of odd-frequency $s$-wave superconductivity is
demonstrated in the two-channel Kondo lattice by means of the dynamical mean-field theory 
combined with the continuous-time quantum Monte Carlo method.
Around half filling of the conduction bands, divergence of an odd-frequency 
pairing susceptibility is found, which signals instability toward the superconductivity. 
The corresponding order parameter is 
equivalent to a staggered composite-pair amplitude with even frequencies, which
involves both localized spins and conduction electrons.
A model wave function is constructed for 
the composite order with use of 
symmetry operations such as charge conjugation and channel rotations.
Given a certain asymmetry of the conduction bands, 
another $s$-wave superconductivity is found that has a uniform order parameter.
The Kondo effect in the presence of two channels is essential for both types of unconventional superconductivity.
\end{abstract}

\pacs{Valid PACS appear here}
\maketitle

\newcommand{\diff}{\mathrm{d}}
\newcommand{\imag}{\mathrm{Im}\,}
\newcommand{\real}{\mathrm{Re}\,}
\newcommand{\trace}{\mathrm{Tr}\,}
\newcommand{\imu}{\mathrm{i}}
\newcommand{\epn}{\mathrm{e}}


Unconventional superconductivity refers to such pairing states that have non-trivial symmetry in spin and/or space-time structures.
Among those 
states, we address the odd-frequency (OF) pairing state \cite{berezinskii74}, 
which breaks the gauge symmetry, but has a zero pairing amplitude at equal time. 
Possible relevance of the OF pairing to real materials
was first 
pointed out by Berezinskii for $^3$He \cite{berezinskii74}.
After the discovery of high-temperature superconductivity in cuprates,
the OF pairing has aroused broad interest 
\cite{emery92, balatsky92, abrahams93, coleman94, zachar96, coleman97, vojta99, fuseya03, yada08, shigeta09, hotta09, shigeta11, kusunose11-2, yanagi12, heid95-2, martisovits98, martisovits00, tanaka12, emery93, balatsky93, coleman93, schrieffer94, coleman95, abrahams95, heid95, jarrell97, anders02, anders02-2, hoshino11, belitz99, solenov09, kusunose11,sakai04, coleman99, flint11, shigeta13}
as one of candidate mechanisms for unconventional superconductivity.

The OF
pairing state can 
be viewed from a different perspective;
it has been recognized that 
the OF superconductivity is alternatively regarded as a composite pairing state with 
even frequencies (EF) \cite{emery92, balatsky93}.
From another argument, it has also been suggested that 
OF superconductivity has a tendency to favor a spatial inhomogeneity \cite{coleman93, heid95},
and that 
finite density of states remains at the chemical potential \cite{balatsky92, coleman93}.

One of possible realizations of OF superconductivity is proposed in the two-channel Kondo systems.
Emery and Kivelson have shown for the two-channel Kondo impurity that 
the OF pairing susceptibility is enhanced at the impurity site \cite{emery92}.
They have further elaborated on a variant of the two-channel Kondo lattice (TCKL) in one dimension \cite{emery93}, and have demonstrated the divergence of the OF pairing susceptibility at zero temperature.
For the TCKL in higher dimensions, microscopic calculations have been performed, however, without finding divergent susceptibility \cite{jarrell97}.
Another calculation for the corresponding Anderson lattice \cite{anders02, anders02-2} did find the 
divergent susceptibility, which vanished as the system goes to the TCKL limit.  
So far no microscopic theory has established the OF pairing in the TCKL
at finite temperature.

In this paper, we present highly accurate numerical results for the pairing susceptibility in high dimensional TCKL, and demonstrate
that the TCKL realizes the $s$-wave OF superconductivity
with the staggered order parameter.
We also derive the corresponding composite order parameter and
model wave functions for the composite orders explicitly, using symmetry arguments such as charge conjugation and channel rotations.
It is further shown, with a certain channel asymmetry, that another $s$-wave superconductivity occurs with EF in the TCKL. 
Our key strategy is to exploit the charge conjugation
that relates the diagonal long-range order to the off-diagonal one.

The TCKL Hamiltonian \cite{jarrell96} is given by $ {\cal H} = {\cal H}_0 + {\cal H}_\mu$ with
\begin{align}
{\cal H}_0 &= \sum_{\bm k\alpha\sigma} \varepsilon_{\bm k}
c_{\bm k\alpha\sigma}^\dagger c_{\bm k\alpha\sigma}
+ J \sum_{i\alpha} \bm S_i \cdot \bm s_{{\rm c}i\alpha}
, \label{eqn_ham1} \\
{\cal H}_\mu&= - \mu \sum_{i\alpha\sigma} 
 c_{i\alpha\sigma}^\dagger c_{i\alpha\sigma}
. \label{eqn_ham2}
\end{align}
The operator $c_{\bm k\alpha\sigma}$ ($c_{i\alpha\sigma}$) annihilates the conduction electron with energy $\varepsilon_{\bm k}$, channel $\alpha=1,2$ and spin $\sigma=\uparrow, \downarrow$ at wave vector $\bm k$ (site $i$).
The spin for conduction electrons $\bm s_{{\rm c}i\alpha} = \tfrac 1 2 \sum_{\sigma\sigma'} c^\dagger_{i\alpha\sigma} \bm \sigma_{\sigma\sigma'} c_{i\alpha\sigma'}$ couple with the local spin $\bm S_i$ antiferromagnetically (i.e. $J>0$).
The chemical potential $\mu$ controls the total number of conduction electrons.
Note that this model has double SU(2) symmetries;
 SU(2)$_{\rm s}$ for spin and SU(2)$_{\rm c}$ for channel degrees of freedom.
We assume 
the bipartite lattice with 
$\varepsilon_{\bm k} + \varepsilon_{\bm k + \bm Q} = 0$, where $\bm Q$ corresponds to a vector at edges of the Brillouin zone.
This assumption makes the model invariant at half filling ($\mu=0$)
under particle-hole 
transformations, {\it i.e.}, 
charge conjugation, 
as will be discussed later.

Physically, the 
degree of freedom described by $\bm S_i$ is interpreted as orbital for the non-Kramers doublet system with $f^2$ configuration in Pr$^{3+}$ and U$^{4+}$ \cite{cox98}.
The labels $\sigma$ and $\alpha$ are then interpreted as orbital and real spin, respectively.
On the other hand, in Kramers systems as in 
Ce$^{3+}$ with $f^1$ configuration, $\bm S_i$ is regarded as a real spin of $f$ electrons.
The labels $\sigma$ and $\alpha$ are then regarded as real spin and orbital, respectively.
The non-Kramers doublet system has the channel (real spin) symmetry protected by the time-reversal symmetry, while in the Kramers doublet case
the channel (orbital) symmetry is only approximate. 
In the following, we call $\sigma$ and $\alpha$ simply as `spin' and `channel', respectively,
unless otherwise stated explicitly.

We use the dynamical mean-field theory (DMFT) \cite{kuramoto85, georges96} for analysis of the TCKL, and the continuous-time quantum Monte Carlo method \cite{rubtsov05, gull11} for the numerical impurity solver.
We take the semi-circular density of states
$\rho (\varepsilon) = (2/\pi) \sqrt{1 - (\varepsilon/D)^2}$
with $D=1$ being the unit of energy.
We have confirmed that the behaviors are qualitatively 
the same if we take the Gaussian density of state for example.

We begin with the pairing susceptibilities in the TCKL.
Following the literature \cite{jarrell97,anders02}
we use for each pairing type the labels C and S indicating `channel' and `spin', 
and s and t indicating `singlet' and `triplet'.
We introduce the following operators dependent on imaginary time to 
describe possible pairings:
\begin{align}
\displaystyle
O_i^{\rm CsSs}(\tau,\tau') &= \sum_{\alpha\alpha'\sigma\sigma'}
c_{i\alpha\sigma} (\tau) 
\epsilon_{\alpha\alpha'} \epsilon_{\sigma\sigma'}
c_{i\alpha'\sigma'} (\tau')
, \label{eq_pair_def1} \\
\displaystyle
O_i^{\rm CsSt}(\tau,\tau') &= \sum_{\alpha\alpha'\sigma}
c_{i\alpha\sigma} (\tau) 
\epsilon_{\alpha\alpha'}
c_{i\alpha'\sigma} (\tau')
, \label{eq_pair_def2} \\
\displaystyle
O_i^{\rm CtSs}(\tau,\tau') &= \sum_{\alpha\sigma\sigma'}
c_{i\alpha\sigma} (\tau) 
\epsilon_{\sigma\sigma'}
c_{i\alpha\sigma'} (\tau')
, \label{eq_pair_def3} \\
\displaystyle
O_i^{\rm CtSt}(\tau,\tau') &= \sum_{\alpha\sigma}
c_{i\alpha\sigma} (\tau) 
c_{i\alpha\sigma} (\tau')
, \label{eq_pair_def4}
\end{align}
where $\epsilon \equiv \imu \sigma^y$ is the antisymmetric unit tensor, and ${\cal A}(\tau) = \epn^{\tau {\cal H}} {\cal A} \epn^{-\tau {\cal H}}$.
Note that $O_i^{\rm CsSs}(\tau,\tau) = O_i^{\rm CtSt}(\tau,\tau) = 0$ due to the Pauli principle,
meaning that these cannot form the 
equal-time
pairings.

In order to calculate susceptibilities, we introduce the two-particle Green function by
$
\chi^{\ell}_{ij} (\tau_1, \tau_2, \tau_3, \tau_4) = \langle T_\tau
O^\ell_i (-\tau_2, -\tau_1)^\dagger O^\ell_j (\tau_3, \tau_4)
\rangle
$
where $T_\tau$ is the time-ordering operator, and $\ell$ represents one of the labels in Eqs.~(\ref{eq_pair_def1}--\ref{eq_pair_def4}).
The Fourier transform is defined by
\begin{align}
\chi^{\ell}_{\bm q} (\imu\varepsilon_n, \imu\varepsilon_{n'})
&= \frac{1}{N\beta^2} \sum_{ij}\int_0^\beta \hspace{-2mm} \diff \tau_1 \cdots \diff\tau_4
\ \chi^{\ell}_{ij} (\tau_1, \tau_2, \tau_3, \tau_4)
\nonumber \\
&\ \ \times 
\epn^{-\imu \bm q \cdot (\bm R_i - \bm R_j)}
\epn^{\imu \varepsilon_n    (\tau_2 - \tau_1)}
\epn^{\imu \varepsilon_{n'} (\tau_4 - \tau_3)}
,
\end{align}
with $\varepsilon_n = (2n+1)\pi T$.
Using this quantity, we define the EF and OF pairing susceptibilities $\chi_{\bm q}^{\ell}$ with 
$\ell\rightarrow \ell_{\rm EF}, \ell_{\rm OF}$
by
\begin{align}
\chi_{\bm q}^{\ell_{\rm EF}} &= \frac{1}{\beta} \sum_{nn'} 
\chi^{\ell_{\rm EF}}_{\bm q} (\imu\varepsilon_n, \imu\varepsilon_{n'}) 
, \label{eq_even_suscep}
\\
\chi_{\bm q}^{\ell_{\rm OF}} &= \frac{1}{\beta} \sum_{nn'} g_n g_{n'} 
\chi^{\ell_{\rm OF}}_{\bm q} (\imu\varepsilon_n, \imu\varepsilon_{n'}) 
, \label{eq_odd_suscep}
\end{align}
where $\ell_{\rm EF}$ denotes CsSt or CtSs, while
$\ell_{\rm OF}$ denotes CsSs or CtSt.
The form factor is defined by $g_n = {\rm sgn }\, \varepsilon_n$ \cite{jarrell97, anders02-2, sakai04}.
Namely we extract the EF part for CsSt and CtSs, and the OF part for CsSs and CtSt.
Equation (\ref{eq_even_suscep}) is the usual susceptibility and must be positive \cite{freericks93}.
On the other hand, the OF susceptibility given by Eq.~(\ref{eq_odd_suscep}) is no longer positive definite due to the presence of 
$g_n$,
but still signals the instability toward the pairing state by its divergence \cite{hoshino11}.
The critical temperature thus obtained is insensitive to
the choice of the form factor provided $g_n$ is odd against $\varepsilon_n$.

\begin{figure}[t]
\begin{center}
\includegraphics[width=75mm]{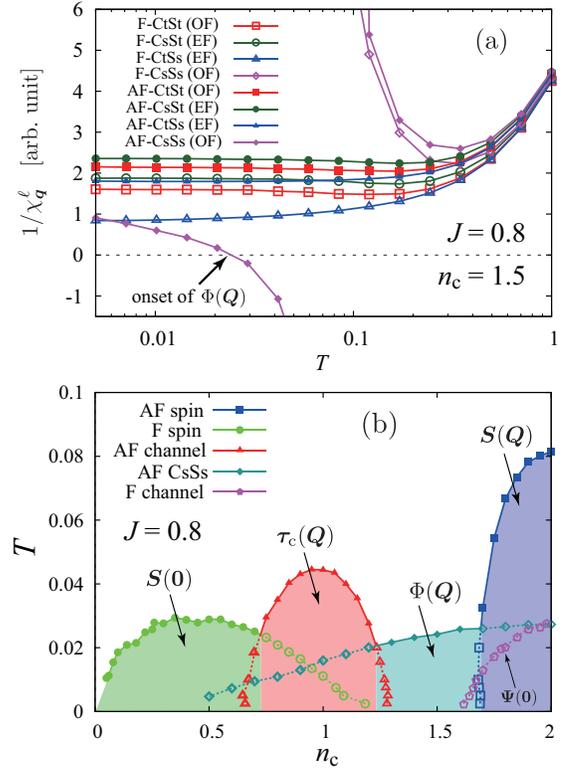}
\caption{
(Color online)
(a) Inverse susceptibilities for EF and OF pairings with uniform (F) or staggered (AF) order.
(b) Phase diagram of the TCKL 
 in the plane of filling ($n_{\rm c}$) and temperature ($T$)
at $J=0.8$.
The dotted lines with blank symbols show referential instability points where another susceptibility has already diverged at higher temperature.
}
\label{fig_phase}
\end{center}
\end{figure}

Figure~\ref{fig_phase}(a) shows the temperature dependence of $\chi^\ell_{\bm q}$ at $J=0.8$ and $n_{\rm c}=1.5$ with $n_{\rm c}$ being the average number of conduction electrons per site.
Here we consider the two ordering vectors $\bm q = \bm 0$ and $\bm q = \bm Q$, which we call ferro (F) and antiferro (AF), respectively.
Among the eight susceptibilities, only the one with AF-CsSs diverges at $T_{\rm sc} \simeq 0.024$ from the negative side, signaling the onset of the OF superconductivity with the ordering vector $\bm Q$.
We have confirmed that similar behaviors are obtained when we take the other parameters such as $J=0.6$ and $J=1.0$.
Since the normal state of the TCKL is a non-Fermi liquid state as seen in the electrical resistivity \cite{jarrell96}, the present system becomes superconducting directly from the non-Fermi liquid.

Together with the diagonal orders that have been obtained in our previous study \cite{hoshino13}, the phase diagram of the TCKL is completed as shown in Fig.~\ref{fig_phase}(b).
Here the diagonal orders are characterized by the vector operators
\begin{align}
& \bm S (\bm q) = \sum_i \bm S_i \epn ^{-\imu \bm q \cdot \bm R_i}
, \\
& \bm \tau_{\rm c} (\bm q) =
\sum_{i\alpha\alpha'\sigma} c^\dagger_{i\alpha\sigma} \bm \sigma_{\alpha\alpha'} c_{i\alpha'\sigma}
\epn ^{-\imu \bm q \cdot \bm R_i}
, \\
& \bm \Psi (\bm q) = \sum_{i\alpha\alpha'\sigma\sigma'}
 c^\dagger_{i\alpha\sigma} \bm \sigma_{\alpha\alpha'} (\bm S_i \cdot \bm \sigma_{\sigma\sigma'}) c_{i\alpha'\sigma'}
\epn ^{-\imu \bm q \cdot \bm R_i}
,
\label{Psi}
\end{align}
which describe spin ($\bm S$), channel ($\bm \tau_{\rm c}$), and composite ($\bm \Psi $) orders, respectively.

The transition temperature to the superconducting AF-CsSs order is lower than the AF spin order at $n_{\rm c}=2$ (half filling) as shown in Fig.~\ref{fig_phase}(b).
As $n_{\rm c}$ decreases, however, the AF-CsSs dominates the AF spin order.
Since the AF-CsSs order can be best visualized at half filling,
we consider the AF-CsSs state mainly at half filling
by neglecting the AF spin order.

We derive the composite order parameter corresponding to the (odd-frequency) AF-CsSs phase by combining the particle-hole and channel-rotation symmetries.
At half filling, the transition temperatures for the AF-CsSs and F channel $[\bm \Psi(\bm 0)]$ orders are the same within the numerical accuracy as seen in Fig.~\ref{fig_phase}(b), which indicates a degeneracy between these two orders.
In fact, these two orders are obtained from each other by symmetry operations
at half filling.
To demonstrate this, we introduce a particle-hole transformation $\mathscr{P}_2$ that acts only on channel $\alpha=2$ as
\begin{align}
\mathscr{P}_2c_{i2\sigma} \mathscr{P}_2^{-1} &= 
\sum_{\sigma'} \epsilon_{\sigma\sigma'} c^\dagger_{i2\sigma'}
\epn^{\imu\bm Q \cdot \bm R_i}.
\label{ph}
\end{align}
On the other hand,
$c_{i1\sigma}$ and $\bm S_i$ are not affected by $\mathscr{P}_2$.
The half-filled Hamiltonian ${\cal H}_0$ and composite quantity $\Psi^z (\bm 0)$ for the F-channel order are invariant under this transformation.
Here the phase factor in Eq.(\ref{ph}) 
is necessary to make the kinetic energy invariant under $\mathscr{P}_2$.

By contrast,
the transverse components $\Psi^\pm (\bm 0) = \Psi^x (\bm 0) \pm \imu \Psi^y (\bm 0)$ in Eq.(\ref{Psi}) are affected by $\mathscr{P}_2$.
The  explicit form of
$\Phi(\bm Q)^\dagger \equiv \mathscr{P}_2 \Psi^+ (\bm 0) \mathscr{P}_2^{-1}$ 
is given by
\begin{align}
\Phi(\bm Q)^\dagger 
= \sum_{i\alpha\alpha'\sigma\sigma'}
c^\dagger_{i\alpha\sigma} \epsilon_{\alpha\alpha'} [\bm S_i \cdot (\bm \sigma\epsilon)_{\sigma\sigma'}] c^\dagger_{i\alpha'\sigma'}
\epn ^{\imu \bm Q \cdot \bm R_i}
. \label{eqn_trans_composite}
\end{align}
This composite quantity gives the EF order parameter corresponding to the AF-CsSs phase.
Thus the AF-CsSs and F-channel orders are exactly degenerate at half filling.
This is also interpreted as a reflection of the SO(5) symmetry at $\mu=0$ \cite{affleck92, hattori12}.
The degeneracy is lifted by the chemical potential as shown in Fig.~\ref{fig_phase}(b).

Another form of the order parameter can be constructed that
involves conduction electrons only.
The simplest derivation is to start again from the F-channel order, and apply 
$\mathscr{P}_2$.
In the F-channel phase with $\Psi^z (\bm 0)$, the conduction electrons with $\alpha=1$ are nearly free, while the ones with $\alpha=2$ form the Kondo insulator at half filling \cite{hoshino11}.
Hence the difference 
$
\sum_{\bm k \sigma} \varepsilon_{\bm k} \langle c^\dagger_{\bm k1\sigma}c_{\bm k 1 \sigma} - c^\dagger_{\bm k2\sigma}c_{\bm k 2 \sigma}\rangle
$ 
of the kinetic energies between channels arises, which
can be regarded as a secondary order parameter \cite{nourafkan08, hoshino13}.
We write the quantity in the SU(2)$_{\rm c}$ symmetric form as
\begin{align}
\bm \psi_{\rm c} (\bm 0) = \sum_{\bm k \alpha \alpha' \sigma} \varepsilon_{\bm k}
c^\dagger_{\bm k\alpha\sigma} \bm \sigma_{\alpha\alpha'} c_{\bm k\alpha'\sigma}
. \label{eqn_kinetic}
\end{align}
Performing the particle-hole transformation, we obtain 
$\phi_{\rm c} (\bm Q)^\dagger \equiv \mathscr{P}_2 \psi^+_{\rm c} (\bm 0) \mathscr{P}_2^{-1}$
as
\begin{align}
\phi_{\rm c} (\bm Q)^\dagger 
= \sum_{\bm k\alpha\alpha'\sigma\sigma'} \varepsilon_{\bm k}
c^\dagger_{\bm k\alpha\sigma} 
\epsilon_{\alpha\alpha'} \epsilon_{\sigma\sigma'}
c^\dagger_{-\bm k-\bm Q, \alpha'\sigma'}
. \label{eqn_second_pair}
\end{align}
Note that this expression is similar to the so-called $\eta$ pairing \cite{yang89}.  However, a difference lies in the form factor $\varepsilon_{\bm k}$ present in Eq.~(\ref{eqn_second_pair}).
Direct calculation shows that
$\Phi (\bm Q)$ and $\phi_{\rm c} (\bm Q)$ are related to the time-derivative
$\partial O_i^{\rm CsSs} (\tau,0) / \partial \tau |_{\tau=0}$
of the OF order given by Eq.~(\ref{eq_pair_def1}).
If one would investigate the instability toward superconductivity using the EF susceptibility for $\Phi(\bm Q)$ or $\phi_{\rm c} (\bm Q)$, explicit calculation of the correlation function would be much more tedious. 
Hence the OF susceptibility defined in Eq.~(\ref{eq_odd_suscep}) provides a convenient tool serving for the same purpose.

\begin{figure}[t]
\begin{center}
\includegraphics[width=80mm]{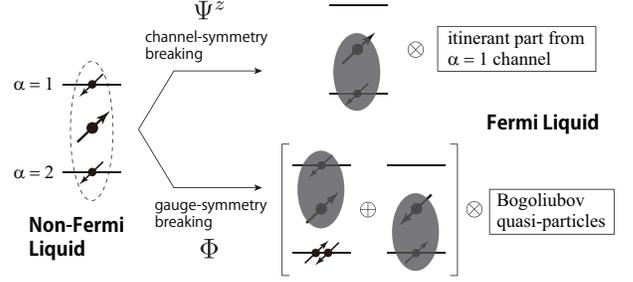}
\caption{
Local configuration of electrons for disordered non-Fermi liquid state (left),
and diagonally ($\Psi^z$) or off-diagonally ($\Phi$) ordered state (right).  The ordered states are described by Eqs.~(\ref{eqn_wf_chan}) and (\ref{eqn_wf_super}).
}
\label{fig_screen}
\end{center}
\end{figure}

Let us proceed to construct wave functions for the ordered phases.
A simple form, though crude, 
should be useful to visualize a composite order.
Starting from the F-channel state at half filling, we obtain a superconducting state by symmetry operations on the wave function.
According to Ref.~\cite{hoshino11}, the F channel state described by 
$\Psi^z(\bm 0) = 2\sum_i \bm S_i \cdot (\bm s_{{\rm c}i1} - \bm s_{{\rm c}i2})$
consists of itinerant electrons for $\alpha=1$ and the Kondo singlets for $\alpha = 2$.
We introduce a simplified wave function describing this F channel state as
\begin{align}
|\Psi^z (\bm 0)\rangle = \prod_{\bm k\in{\rm HBZ},\sigma} c^\dagger_{\bm k1\sigma}
\prod_{i}|{\rm KS}\rangle_{i2}
, \label{eqn_wf_chan}
\end{align}
where `HBZ' means a half Brillouin zone.
The $\alpha=1$ part labeled by $\bm k$ represents free conduction electrons at half filling.
The local Kondo singlet state with channel $\alpha$ is written as
by 
$|{\rm KS}\rangle_{i\alpha} = (c^\dagger_{i\alpha\uparrow} |\downarrow\rangle_i - c^\dagger_{i\alpha\downarrow} |\uparrow \rangle_i )/\sqrt 2$ with $|\sigma \rangle_i$
being the localized-spin state at site $i$.
The local picture of Eq.~(\ref{eqn_wf_chan}) is illustrated in the upper-right part of Fig.~\ref{fig_screen}.

To obtain the off-diagonal order, we combine $\mathscr{P}_2$ with 
a channel mixing unitary transformation $\mathscr{R}$ defined by
\begin{align}
\mathscr{R}c_{i1(2)\sigma} \mathscr{R}^{-1} &= 
\left[ c_{i1\sigma} +(-) c_{i2\sigma} \right] / \sqrt 2
,
\end{align}
which rotates from $z$ to $x$
axis in 
channel space.
In view of 
$\mathscr{P}_2\mathscr{R} {\cal H}_0 (\mathscr{P}_2\mathscr{R})^{-1}= {\cal H}_0$
and
$\mathscr{P}_2\mathscr{R} {\Psi^z} (\bm 0) (\mathscr{P}_2\mathscr{R})^{-1}= \frac 1 2 [\Phi (\bm Q) + \Phi (\bm Q)^\dagger]$,
it is reasonable to postulate
a model wave function $|\Phi (\bm Q) \rangle \equiv \mathscr{P}_2\mathscr{R}|\Psi^z (\bm 0) \rangle$
for the AF-CsSs state as 
\begin{align}
&|\Phi (\bm Q) \rangle = \prod_{\bm k\in{\rm HBZ},\sigma} \frac{1}{\sqrt 2}
\left(
c^\dagger_{\bm k1\sigma} + 
\sum_{\sigma'} \epsilon_{\sigma\sigma'} c_{-\bm k-\bm Q, 2\sigma'} \right)
\nonumber \\
&\ \ \ \ \times
\prod_{i} \frac{1}{\sqrt 2}
\left(
 c^\dagger_{i2\uparrow}c^\dagger_{i2\downarrow}|{\rm KS}\rangle_{i1}
+
\epn^{\imu \bm Q \cdot \bm R_i} |{\rm KS}\rangle_{i2}
\right)
. \label{eqn_wf_super}
\end{align}
Note that $\mathscr{P}_2$ transforms the vacant state into the states occupied doubly by electrons with channel $\alpha=2$ at each site.
As seen in the second line of Eq.~(\ref{eqn_wf_super}), the states with one and three local conduction electrons per site are superposed, indicating  the broken gauge symmetry.
As is clear from this expression, both channels $\alpha=1, 2$ participate to make the local spin-singlet state. 
The lower-right part of Fig.~\ref{fig_screen} illustrates this local state.
The first line of Eq.~(\ref{eqn_wf_super}) includes 
the Bogoliubov quasi-particles 
composed of particle at $\bm k$ and hole at $-\bm k-\bm Q$, which in general have
a finite density of states at the Fermi level.

We now consider the region around $n_{\rm c}=1$ in Fig.\ref{fig_phase}(b) where
the AF-channel order $\bm \tau_{\rm c} (\bm Q)$ is dominant.
By the particle-hole transformation $\mathscr{P}_2$,
we can relate the electronic state at $n_{\rm c}=1$ to the half-filled case 
under asymmetric channel potential.
We introduce the new Hamiltonian
\begin{align}
\tilde {\cal H} \equiv \mathscr{P}_2{\cal H} \mathscr{P}_2^{-1} = {\cal H}_0 -\mu \, \tau_{\rm c}^z (\bm 0)
.
\end{align}
Starting from $n_{\rm c}=1$ for $\cal H$, we end up in $\tilde{\cal H}$ with 
1/2 electron per site for $\alpha=1$, while 3/2 electrons 
for $\alpha=2$.
Then we obtain the sum $\tilde{n}_{\rm c}=2$.
Namely the hole-doped TCKL described by ${\cal H}$ is transformed into the TCKL at half filling 
under the channel field. 
Physically, $\tilde {\cal H}$ simulates such systems that have two conduction electrons per site but with inequivalent conduction bands.
This situation may arise in a Kondo lattice with Kramers degeneracy, since 
the channels in this case correspond to spatial orbitals that have no degeneracy in general.

On the other hand,
the AF-channel order 
at $n_{\rm c} = 1$ 
are transformed as
\begin{align}
\mathscr{P}_2 \tau_{\rm c}^z (\bm Q) \mathscr{P}_2^{-1} &= 
\sum_{i\alpha\sigma} c^\dagger_{i\alpha\sigma}  c_{i \alpha\sigma}
\epn^{\imu\bm Q \cdot \bm R_i}
, \label{eqn_cdw}
\\
\mathscr{P}_2 \tau_{\rm c}^+ (\bm Q) \mathscr{P}_2^{-1} &= 
\sum_{i\alpha\sigma\sigma'} c^\dagger_{i\alpha\sigma} \epsilon _{\sigma\sigma'} c^\dagger_{i \alpha\sigma'}
\equiv p_{\rm c}(\bm 0)^\dagger
. \label{eqn_s_wave_sc}
\end{align}
Namely, the AF-channel order in ${\cal H}$ corresponds to the charge density wave 
given by Eq.~(\ref{eqn_cdw}),
as well as
the $s$-wave superconductivity $p_{\rm c}(\bm 0)$ given by Eq.~(\ref{eqn_s_wave_sc})
in the new Hamiltonian $\tilde {\cal H}$.
These two orders are degenerate at $\tilde{n}_{\rm c}=2$ by symmetry.
We have numerically confirmed that this degeneracy is lifted with $\tilde{n}_{\rm c}\neq 2$, 
and the $s$-wave superconductivity $p_{\rm c} (\bm 0)$ is more stable.
This $s$-wave pairing can be understood from the strong coupling limit,  
following interpretation of the AF channel order in ${\cal H}$ \cite{cox98, schauerte05}. 
Then, the local
image of the $p_{\rm c}(\bm 0)$ state is given by the lower-right panel of Fig.~\ref{fig_screen} without the 
Bogoliubov quasi-particle part.
In contrast to the AF-CsSs state, 
the superconductivity with $p_{\rm c}(\bm 0)$ has a full gap in the density of states.

In a similar manner, we can show that
the off-diagonal AF-CsSs order in ${\cal H}$ is transformed into the diagonal composite order 
$\Psi^{\pm} (\bm 0)$ in $\tilde {\cal H}$.
The spin order $\bm S (\bm q)$ remains the same after the transformation.
Thus the phase diagram for $\tilde {\cal H}$ is obtained without further calculation
if we replace $n_{\rm c}$ by $\langle \tau_{\rm c}^z(\bm 0)\rangle$ in Fig.~\ref{fig_phase}(b).

Finally let us briefly discuss possible relevance of our results to real physical systems.
The primary candidate for the non-Kramers doublet system 
with the order $\Phi (\bm Q)$ 
is UBe$_{13}$, which has first been proposed as the two-channel Kondo system by Cox \cite{cox87}.
The superconductivity in UBe$_{13}$ appears directly from the non-Fermi liquid state \cite{ott83}, which is consistent with what we have found in the TCKL.
For Kramers-doublet systems, on the other hand, a channel (orbital) asymmetry is inevitable.
Then the modified Hamiltonian $\tilde {\cal H}$ seems relevant to describe the two-orbital Kondo lattice.
Recent specific-heat measurement in CeCu$_2$Si$_2$ \cite{kittaka13}
has reported a full-gap superconductivity reminiscent of multiband superconductivity.
Hence we suggest possible relevance of the $p_{\rm c}(\bm 0)$ state to CeCu$_2$Si$_2$.

In conclusion, taking the TCKL, 
we have demonstrated instability of the non-Fermi liquid state toward
the OF $s$-wave superconductivity with finite center-of-mass momentum,
 and with channel-singlet and spin-singlet pairing.
The OF pairing state is alternatively characterized by an EF composite order. 
We have further 
derived another EF $s$-wave superconductivity with the uniform order parameter by considering asymmetry in the channels.
Further studies inside the superconducting phases will provide a better understanding of the unconventional superconductivity.

We are grateful to Yusuke Kato for stimulating discussions and valuable comments on our paper.
We also appreciate Kazumasa Hattori, Hiroaki Kusunose
and Youichi Yanase for fruitful discussions.
S.H. acknowledges the financial support from Japan Society for Promotion of Science.

%
%
%

\end{document}